\documentstyle[aps,12pt]{revtex}
\input epsf
\begin{document}
\title{{\normalsize{{\hskip 8cm} BIHEP-TH-97-17  , Nov. 1997}}\\
Momentum Spectrum of $\eta'$ Meson in the Inclusive
Decay $B\to \eta'X_s$}
   
\author{Dong-Sheng Du and Mao-Zhi Yang\\
 CCAST(WORLD LABORATORY)P.O.BOX 8730,BEIJING,100080\\
                               and \\
   Institute of High Energy Physics, Chinese Academy of Sciences,\\
   P.O.Box 918(4), Beijing, 100039, People's Republic of China
\thanks{Mailing address}}
\maketitle
\begin{center}
PACS numbers: 13.25.Hw, 13.40.Hq 
\end{center}	     

\vskip 0.3in
\begin{abstract}
The momentum distribution for the $\eta'$ meson produced through the inclusive
decay $B\to \eta'X$ is investigated under two decay mechanisms, 
$b\to \eta's$ and $b\to \eta' s g$. Although all these two mechanisms 
can explain the recently
observed decay rate of $B\to \eta'X$, the momentum spectrums for $\eta'$
meson predicted by them are strongly different. Thus detailed experiment is
proposed to distinguish the two cases.
\end{abstract}
\newpage

Recently more attention has been focused on the inclusive and exclusive decay
mode of B mesons, $B\to \eta'X_s$ and $B\to \eta'K$, which have been observed
with relatively large decay rates \cite{1}, 
$Br(B\to \eta'X_s)=(6.1\pm1.6\pm1.3)\times 10^{-4}$ and 
$Br(B\to \eta'K)=(7.1^{+2.5}_{-2.1}\pm0.9)\times 10^{-5}$, under the 
constraint: $2.2GeV\leq E(\eta')\leq 2.7GeV$. Many authors have made efforts
to explain these conspicuously large decay rates \cite{2,3,add}. The mechanism via the 
intermediate formation of a $c\bar{c}$ pair to form the final state containing
an $\eta'$ meson is too weak to explain the observed large decay rate \cite{2}.
Besides, there are also mechanisms based on two-boby decay of b-quark
$b\to s\eta'$\cite{2} and three-body decay $b\to sg\eta'$\cite{3}. In the 
two-body decay mechanism, the b-quark decays into s-quark under gluon emission, 
where the gluon is either on-shell or off-shell, then the gluon transformates
into $\eta'$ meson due to the anomalously 
large coupling of $\eta'$-gluon system
\cite{4}. This is a highly non-perturbative process. While in the three-body
decay mechanism, the process proceeds via $b\to sg^*$ with $g^*$ transfers into
$g$ and $\eta'$ because of the anomalously large $\eta'-g^*-g$ coupling\cite{3}. All
these two mechanisms can explain the experimental results with suitable 
coupling constant of $\eta'$ and gluon field.

In this work we want to study the momentum distribution 
of $\eta'$ meson based on 
the  two-body and three-body decay mechanisms $b\to s\eta'$ and $b\to sg\eta'$.
By comparing the momentum distributions of $\eta'$, the two mechanisms can be
distinguished in experiment if the momentum spectrum of $\eta'$ is measured. 

For the two-body decay mechanism, Fritzsch has given an ansatz for the 
effective interaction between the quarks and the gluonic densities\cite{2},
$$ {\cal H}^{eff}=-const.\frac{G_F}{\sqrt{2}} V_{ts}\frac{\alpha_s}{4\pi^2}
                \cdot m_b\bar{b}_Rs_L  (G_{\mu\nu}G^{\mu\nu}+
                  G_{\mu\nu}\tilde{G}^{\mu\nu}), \eqno(1)$$
where the const. includes the highly non-perturbative contributions of the 
strong interactions involved in the decay process. Using the factorization
method the amplitudes of $B\to\eta' X_s$ can be calculated,
$$\langle X\eta'|{\cal H}^{eff}|B\rangle = -f^{\eta'}_{g^*}
                    \frac{G_F}{\sqrt{2}} V_{ts}\frac{\alpha_s}{4\pi^2}   
                    m_b\bar{b}_Rs_L, \eqno(2)$$
where $f^{\eta'}_{g^*}\equiv const.\langle\eta'|G_{\mu\nu}G^{\mu\nu}+
                     G_{\mu\nu}\tilde{G}^{\mu\nu})|0\rangle$. Here no attempt
is made to calculate $f^{\eta'}_{g^*}$, because of its strong non-perturbative
origin, but to fit the experimental data to get it.
With eq.(2), the decay width is obtained,
$$\Gamma_0=\frac{1}{8\pi}|f^{\eta'}_{g^*}|^2\frac{G_F^2\alpha_s^2|V_{ts}|^2}
            {128\pi^2} |k_c|, \eqno(3)$$
where $|k_c|=\frac{1}{2m_b}[(m_b^2-(m_s+M_{\eta'})^2)(m_b^2-(m_s-M_{\eta'})^2 )]
     ^{1/2}$. 
     
     To get the momentum distribution the Fermi motion of the spectator quark
inside the B meson should be taken into account. In this work the Fermi 
motion is considered using the same method in ACCMM model\cite{5}. 
The spectator quark inside the 
meson is handled as an on-shell particle, and is attributed with a Fermi 
motion. Thus the spectator quark has definite mass $m_{sp}$ and its momentum
is $|\vec{p}|=p$. The b quark is considered to be off-shell, its virtual mass
$M(p)$ can be given in the restframe of the B meson due to the four-momentum
conservation as
$$M(p)=\left(M_B^2+m_{sp}^2-2M_B\sqrt{m_{sp}^2+p^2}~\right)^{1/2}. \eqno(4)$$
The momentum distribution probability of the Fermi motion $\phi (p)$ is 
introduced with a Gaussian behaviour in the ACCMM model,
$$\phi (p)=\frac{4}{\sqrt{\pi}p_F^3} e^{-p^2/p_F^2}, \eqno(5)$$
where $\frac{4}{\sqrt{\pi}p_F^3}$ is the normalizing constant under the 
normalization $\int^\infty_0dp p^2\phi(p)=1$. The free parameter $p_F$ is
taken to be $p_F=0.57GeV$, which has been used to explain the momentum spectrum
of $J/\psi$ in $B\to J/\psi X$ \cite{6}.

The decay width of b quark should be transformed from the b quark 
restframe to the B meson restframe with a Lorentz boost,
$$\frac{d\Gamma_b(p_{\eta'},p)}{d p_{\eta'}}=\left\{ 
   \begin{array}{ll}
   0, & p_{\eta'}<|k_-|,\\
   \displaystyle\frac{M_b}{E_b}\frac{\Gamma_0}{k_+-|k_-|}, & 
                  |k_-|<p_{\eta'}<k_+,\\
   0, &p_{\eta'}>k_+
   \end{array}\right. .  \eqno(6)$$
where $\Gamma_0$ is expressed in eq.(3) with $m_b$ be substituted with $M(p)$,
and $k_\pm=\frac{1}{M_b}(E_bp_0\pm E_0 p)$, $E_0$, $p_0$ are the energy and 
momentum of $\eta'$ meson in the restframe of b quark, respectively. To compare
the momentum distribution with experiment, one has to transform eq.(6) into
the laboratory frame where the B meson is in flight. Because the B mesons
are produced through the cascade process $e^+e^-\to \Upsilon (4s)\to B\bar{B}$,
they move with a momentum $p_B=0.34 GeV$. Performing the Lorantz boost to the
laboratory frame, the differential branching ratio for a B meson in flight is,
$$\frac{1}{\Gamma_B}\frac{d\Gamma}{d p_{\eta'}}=\tau_B 
   \int\limits^{\hat{K}_+(p_{\eta'})}_{|K_-(p_{\eta'})|} \frac{dp'_{\eta'}}
   {k_+(p'_{\eta'})-|k_-(p'_{\eta'})|}\int^{p_{max}}_0 dp p^2\phi(p) 
   \frac{d\Gamma_b}{dp'_{\eta'}}(p'_{\eta'},p),  \eqno(7)$$
where $p_{max}=\frac{1}{2M_B}[(M_B^2+m_{sp}^2-
   M_{\eta'}^2)^2-4M_B^2m_{sp}^2]^{1/2}$, and 
$$K_\pm(k)=\frac{1}{M_B}(E_Bk\pm p_BE_{\eta'}), ~~~~~
\hat{K}_+(k)=min\{K_+(k),k_{max}\} , \eqno(8)$$
$k_{max}$ is the maximum value of $\eta'$ momentum in the restframe of B meson.

The numerical result is shown by the solid curve of Fig.1. The discussion is 
given in the end part of the paper.

\begin{figure}
\centerline{\epsfysize = 5 in \epsffile {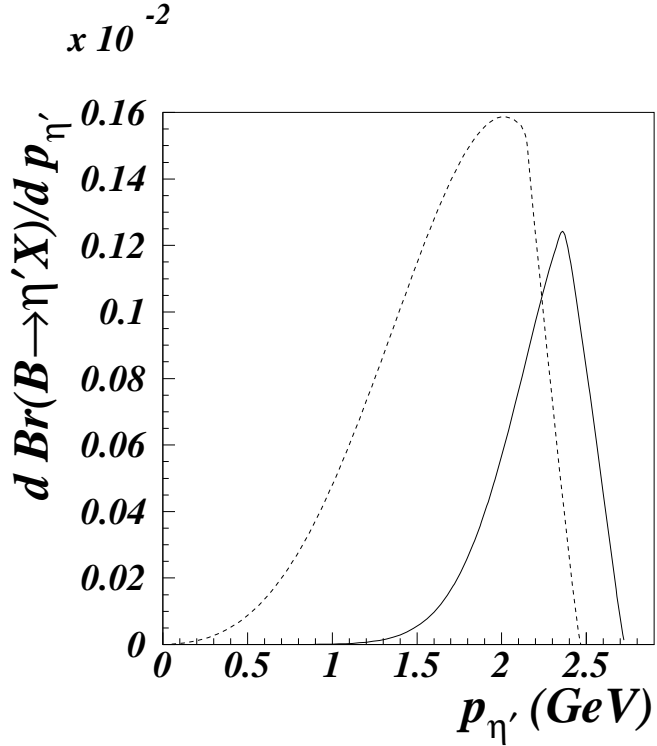}}
\caption{Momentum distribution of $\eta'$ meson in the decay of $B\to\eta' X$
 in the laboratory frame where B meson is in flight. The solid curve is for
 the $b\to s\eta'$ case, while the dashed curve for the $b\to sg\eta'$ case.
}.
\end{figure}

Now we turn to the three-body decay mechanism $b\to sg\eta'$, for the detail
of which the readers are referred to Ref.\cite{3}. The loop induced $b\to s$
current in the standard model is \cite{7}
$$\frac{G_F}{\sqrt{2}} \frac{g_s}{4\pi^2} v_{t}\bar{s}t^a\{\Delta F_1(q^2
  \gamma_\mu-q_\mu q{\hskip -1.8mm/} )L-
  F_2 i \sigma_{\mu\nu}q^\nu m_b R\}b, \eqno(9)$$
where $v_t=V^*_{ts}V_{tb}$, $\Delta F_1=-5.05$, $F_2\simeq 0.2$. The Feynman 
rule for the $\eta'-g-g$ vertex is \cite{3},
$$-i a_g c_p 
 \epsilon_{\mu\nu\alpha\beta}\epsilon^\mu(q)\epsilon^\nu(k) q^\alpha
 k^\beta, \eqno(10)$$
here $a_g$ is the effective gluon anomaly coupling. $\epsilon$ is the 
polization of the gluon field, $q$ and $k$ are the four-momentum of the two
gluons. 

Define $x\equiv \frac{p_x^2}{m_b^2}$, $y\equiv\frac{q^2}{m_b^2}$, 
$x' \equiv \frac{M_{\eta'}^2}{m_b^2}$, $x'_s \equiv \frac{m_s^2}{m_b^2}$, $p_x$
is the four-momentum of all the inclusive hadranic products except $\eta'$ 
meson. The differential branching ratio is 
$$\frac{1}{\Gamma_B}\frac{d\Gamma (b\to sg\eta')}{d p_{\eta'}}=
  \frac{1}{\Gamma_B} \frac{G^2_F|v_t|^2m_b^5}{256\pi^3} \frac{g_s^2}{16\pi^4}
  \frac{a_g^2}{4} \int dy [\Delta F_1^2 C_1+F_2\Delta F_1 C_2+F_2^2 C_3]
  \frac{2 m_b p_{\eta'}}{E_{\eta'}}, \eqno(11)$$
with $C_0=[-2(x-x_s')^2y+(1-y-x_s')(y-x')(2x+y-x'-2x_s')]/2$,
$C_1=-(1-y-x_s')(y-x')^2/y$, and $C_2=[2(x-x_s')^2y^2-(1-y-x_s')(y-x')
   (2(x-x'_s)y-(1-x_s')(y-x'))]/2y^2$, $x=1+x'-2\sqrt{x'+p_{\eta'}^2/m_b^2}$.
    $C_0$, $C_1$, $C_2$ are consistent with
   these of Ref.\cite{3} (Hou and Tseng) when taking the limit $x_s'\to 0$.
   
Fot the numerical calculation, we take $m_b=4.8 GeV$, $m_s=0.175GeV$, and 
$a_g=1.70GeV^{-1}$ which can ensure that the integrated branching ratio under the
constraint $2.2GeV\leq E_{\eta'} \leq 2.7 GeV$ is the center value of the 
experimental
data. Using eq.(7) to transform eq.(11) into the laboratory frame with B meson
in flight, the result of the differential branching ratio is shown by the
dashed curve of Fig.1.

Fig.1 shows that in the three-body case (dashed curve), the momentum
distribution for $\eta'$ meson is much broader than that in the two-body decay
case (the solid curve). At the same time most part of the curve lies lower
than the cut used in the experiment $2.2GeV\leq E_{\eta'}\leq 2.7GeV$
$(2.0 GeV\leq p_{\eta'}\leq 2.52 GeV)$. While in the $b\to \eta's$ case, no
$\eta'$ meson is found with momentum less than $1.5GeV$. The two mechanisms
can be distinguished by experiment. Thus more detailed experiment is needed.

\vspace{2cm}

\section*{Acknowledgements}

This work was supported in part by the China National Nature Science
Foundation and the Grant of State Commission of Science and Technology
of China.

\newpage

 
\end{document}